\documentclass[sigconf]{acmart}

\usepackage{balance}

\AtBeginDocument{%
  }

\copyrightyear{2025}
\acmYear{2025}
\setcopyright{cc}
\setcctype{by}\setcctype{by}
\acmConference[CIKM '25]{Proceedings of the 34th ACM International
Conference on Information and Knowledge Management}{November 10--14,
2025}{Seoul, Republic of Korea}
\acmBooktitle{Proceedings of the 34th ACM International Conference on
Information and Knowledge Management (CIKM '25), November 10--14, 2025,
Seoul, Republic of Korea}\acmDOI{10.1145/3746252.3760942}
\acmISBN{979-8-4007-2040-6/2025/11}

\usepackage{mathrsfs}
\usepackage{appendix}
\usepackage{balance}
\usepackage{colortbl}
\usepackage{amsmath}
\usepackage{booktabs}
\usepackage{tabularx}
\usepackage{multirow}
\usepackage{caption}
\usepackage{subcaption}
\usepackage{adjustbox}

\usepackage[T1]{fontenc}

\usepackage{hyperref}
\usepackage{enumitem}

\DeclareMathOperator*{\argmax}{arg\,max}

\usepackage{tcolorbox}

\usepackage{url}
\usepackage{calrsfs}

\usepackage{soul}
\usepackage{tikz}

\newcommand{\para}[1]{\paragraph{\textnormal{\textbf{#1}}}}

\usepackage{colortbl}
\definecolor{lightblue}{RGB}{173, 216, 230}

\definecolor{mygray}{gray}{0.9} 

\newcommand{\gc}[1]{\cellcolor{mygray}#1}

\usepackage{adjustbox}

\usepackage{fix-cm}
\DeclareMathAlphabet{\pazocal}{OMS}{zplm}{m}{n}
\usepackage{arydshln}

\usepackage{enumitem} 
\newcommand{\uls}{\begin{itemize}[leftmargin=*]}
\newcommand{\ule}{\end{itemize}}
\newcommand{\ols}{\begin{enumerate}[leftmargin=*]}
\newcommand{\ole}{\end{enumerate}}
\newcommand{\li}{\item}

\usepackage[ruled,vlined,linesnumberedhidden]{algorithm2e}

\renewcommand{\vec}[1]{\mathbf{#1}}

\usepackage{arydshln}
\newcommand{\icln}{L-RAG}

\newcommand{\iclz}{0-shot}
\newcommand{\rag}{U-RAG}
\newcommand{\moe}{LU-RAG}

\newcommand{\rob}{RoBERTa}
\newcommand{\hfrag}{HF-RAG}
\newcommand{\lora}{LoRA}

\usepackage{diagbox}

\begin{document}

\title{HF-RAG: Hierarchical Fusion-based RAG with Multiple Sources and Rankers}

\author{Payel Santra}
\orcid{0009-0005-5721-248X}
\authornote{Both authors contributed equally to this research.}
\affiliation{%
\institution{IACS}
  \city{Kolkata}
  \country{India}
}
\email{payel.iacs@gmail.com}

\author{Madhusudan Ghosh}
\orcid{000-0002-8330-2703}
\authornotemark[1]
\affiliation{%
\institution{IACS}
  \city{Kolkata}
  \country{India}
}
\email{madhusuda.iacs@gmail.com}

\author{Debasis Ganguly}
\orcid{0000-0003-0050-7138}
\affiliation{%
  \institution{University of Glasgow}
  \city{Glasgow}
  \country{United Kingdom}
}
\email{debasis.ganguly@glasgow.ac.uk}

\author{Partha Basuchowdhuri}
\orcid{0000-0001-7655-7591}
\affiliation{%
\institution{IACS}
  \city{Kolkata}
  \country{India}
}
\email{partha.basuchowdhuri@iacs.res.in}

\author{Sudip Kumar Naskar}
\orcid{0000-0003-1588-4665}
\affiliation{%
  \institution{Jadavpur University}
  \city{Kolkata}
  \country{India}
}
\email{sudip.naskar@gmail.com}

\renewcommand{\shortauthors}{Santra et al.}

\begin{abstract}
Leveraging both labeled (input-output associations) and unlabeled data (wider contextual grounding) may provide complementary benefits in retrieval augmented generation (RAG). However, effectively combining evidence from these heterogeneous sources is challenging as the respective similarity scores are not inter-comparable. Additionally, aggregating beliefs from the outputs of multiple rankers can improve the effectiveness of RAG. Our proposed method first aggregates the top-documents from a number of IR models using a standard rank fusion technique for each source (labeled and unlabeled). Next, we standardize the retrieval score distributions within each source by applying z-score transformation before merging the top-retrieved documents from the two sources. We evaluate our approach on the fact verification task, demonstrating that it consistently improves over the best-performing individual ranker or source and also shows better out-of-domain generalization.
\end{abstract}




\maketitle

\section{Introduction} \label{sec:intro}

\begin{figure}[t]%
\centering
{\includegraphics[width=0.37\textwidth]{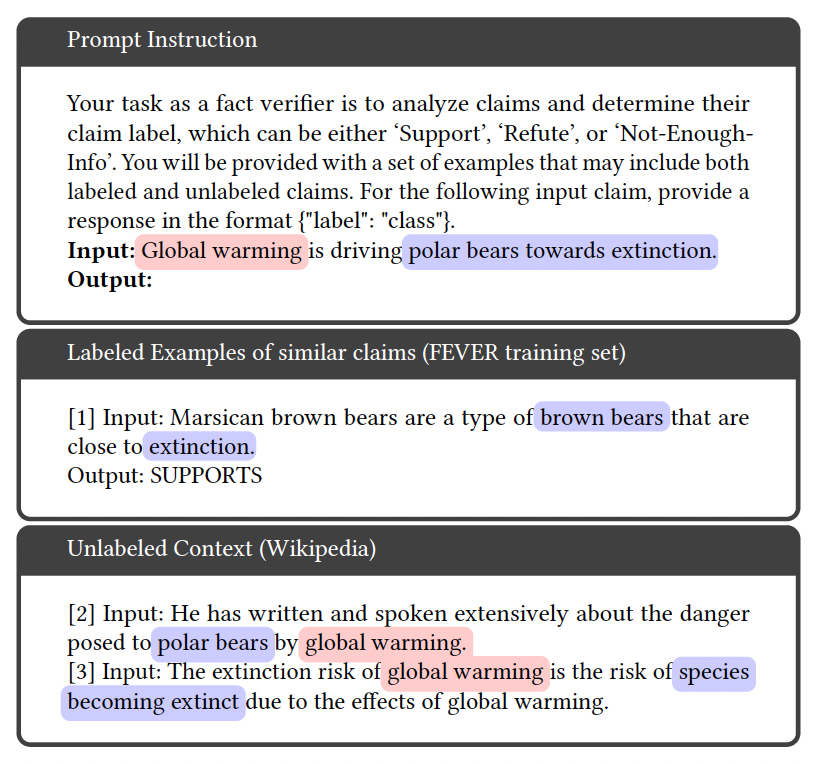}}
\caption{Our proposed approach {\hfrag} leverages both labeled and unlabeled data to provide sub-topic–specific contextual information.}
\label{fig:prompt_template}
\end{figure}

While social media platforms enable individuals to access, contribute to, and disseminate information, they also facilitate the rapid and widespread propagation of misinformation and fake news~\cite{kumar2016disinformation,chen2015social}. As such, computational models for automated fact checking, i.e., methods to automatically examine the veracity of claims by retrieving and analyzing supporting or refuting evidence~\cite{schuster2019towards,schuster2021get,jiang2021exploring,asai2023retrieval},
are of high practical importance.
Fact verification approaches include supervised fine-tuning (SFT), in-context learning (ICL), and retrieval augmented generation (RAG). SFT adapts model parameters with labeled data for task-specific learning, while instead of updating model parameters
ICL leverages labeled exemplars to control predictions~\cite{brown2020language, parry2024context} and
RAG includes relevant contextual information from external unlabeled corpora~\cite{lewis2020retrieval, izacard-grave-2021-leveraging, santra2024absence}.

We hypothesize that for fact verification, both labeled and unlabeled data may serve as complementary sources of information, each providing potentially relevant context for different aspects or sub-topics of an input claim. Figure~\ref{fig:prompt_template} illustrates this using a sample claim from the Climate-FEVER dataset: \textit{Global warming causing extinction of polar bears}. In this example, one sub-topic (red highlight) pertains to the adverse effects of global warming, while the other (blue highlight) concerns species extinction more broadly, not limited to polar bears.
The first retrieved example in Figure~\ref{fig:prompt_template}, sourced from the labeled FEVER training set, presents evidence suggesting that brown bears are nearing extinction. Although this does not directly confirm the claim, it supports a plausible hypothesis that polar bears might face a similar threat. This hypothesis is further strengthened by additional contextual information retrieved from Wikipedia, which provides relevant (unlabeled) evidence regarding the broader risks posed by global warming~\cite{santra2025curious, chandra2025one}.

\para{Novel Contributions}
First,
we propose to
\textit{combine information from two distinct sources}--labeled and unlabeled data--to jointly capture both the topic-specific likelihood of a claim being true or false, and the broader contextual information relevant to the input claim.
Second, we propose that
rather than relying on a single ranking model to retrieve topically relevant labeled or unlabeled examples, it is potentially more effective to \textit{aggregate the outputs of multiple rankers}. This approach--commonly used in IR to improve performance~\cite{fox1994combination,farah2007outranking,cormack2009reciprocal}--allows for the fusion of diverse ranking signals.
An overview of our proposed method, which involves a hierarchical combination strategy--first performing intra-ranker fusion within each source, followed by inter-source fusion--is presented in Figure~\ref{main_fig}. Based on this hierarchical fusion mechanism, we refer this approach as \textbf{Hierarchical Fusion-based RAG} (\textbf{HF-RAG}).

\section{Proposed Hierarchical Fusion-based RAG} \label{sec:overview}

\para{Combining Labeled and Unlabeled Contexts in RAG}

Generally speaking, both RAG and ICL can be viewed as mechanisms for incorporating additional contextual information, the former relying on unlabeled documents retrieved from a corpus, while the latter utilizing labeled instances from a training dataset. As a consistent naming convention towards unifying these perspectives, we refer to the former as \textit{Unlabeled RAG} (\textbf{U-RAG}) and the latter as \textit{Labeled RAG} (\textbf{L-RAG}). 
In our proposed approach, we integrate both sources of contextual information -- unlabeled documents and labeled examples -- to leverage their complementary strengths of topical relevance, and task-specific semantics, respectively.
We hypothesize that such a combined approach is
likely to generalize better to new domains, likely because while L-RAG provides the necessary grounding to capture task-specific semantics (input-label associations) required for effective predictions, the inclusion of U-RAG prevents too much overfitting on a particular task itself by capturing a broader task-agnostic semantics thus potentially enabling better generalization to new domains and tasks.

\begin{figure}[t]%
\centering
{\includegraphics[width=0.45\textwidth]{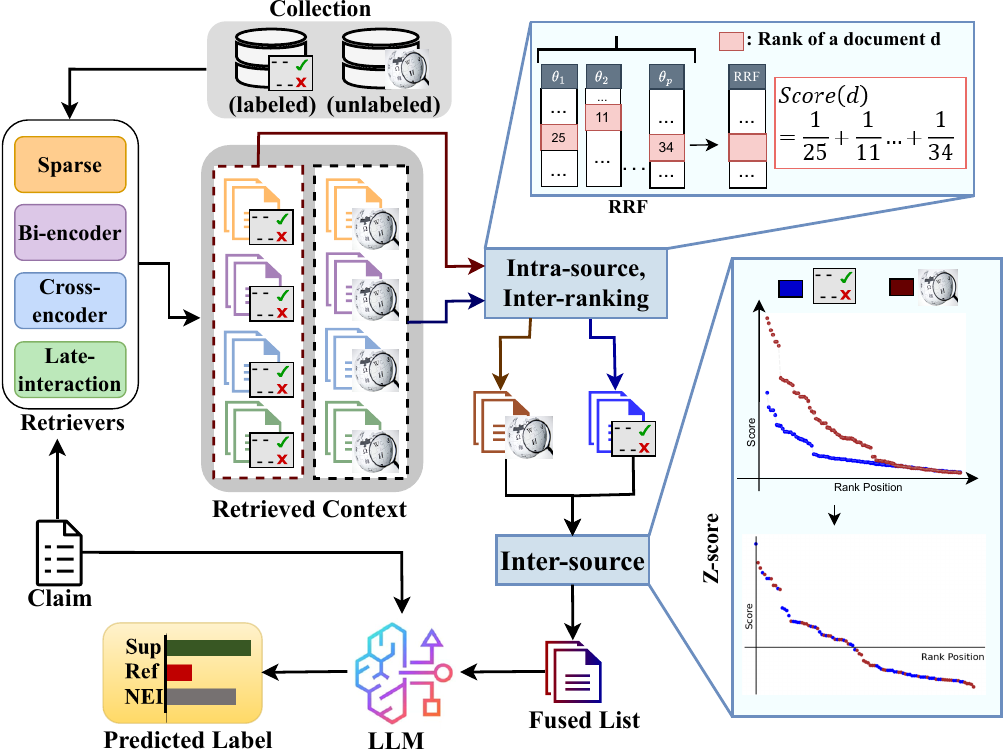}}
\caption{
Schematic overview of our proposed method \hfrag. For a given claim, multiple retrievers are employed to obtain top-ranked documents from labeled and unlabeled sources. These top-documents for each source are combined via reciprocal rank fusion (RRF). These fused lists of non-overlapping documents from the two sources are then merged with a z-score transformation. 
}
\label{main_fig}
\end{figure}

\para{Intra-Source Inter-Ranker Combinations by RRF}

For a specific source (labeled or unlabeled) $C$, an input claim $\vec{x}$, and each IR model $\theta \in \Theta $ (where, $\Theta$ is the set of retrievers) is first invoked to obtain a top-$k$ list of documents $L^{C,\theta}_k$.
Next, we merge each of these top-$k$ lists obtained from each ranker into a single ranked list by the reciprocal rank fusion (RRF)~\cite{cormack2009reciprocal} technique -- a standard fusion method in IR, which computes the overall score of a document as its aggregated reciprocal ranks across each ranked list. Formally, 
\begin{equation}
L^C_k =
\argmax_k \{\overline{\theta_C}(d): d \in \bigcup_{\theta \in \Theta}L^{C,\theta}_k\},\,
\overline{\theta_C}(d) = \sum_{\theta \in \Theta} \frac{1}{\mathrm{rank}(L^{C,\theta}_k, d)},
\label{eq:rrf}
\end{equation}
where $\argmax_k$ denotes a selection of the top-$k$ documents with the highest $\overline{\theta_C}(d)$ scores, $\overline{\theta_C}(d)$ denotes the RRF scores from source $C$, and $\mathrm{rank}(L^{C,\theta}_k, d)$ denotes the rank of a document $d$ in the list $L^{C,\theta}_k$; if $d \notin L^{C,\theta}_k$ then $\mathrm{rank}(L^{C,\theta}_k, d)$ is set to a large number $M(\gg k)$.

Equation \ref{eq:rrf} is applied for each source, $C \in \{l, u\}$ (labeled and unlabeled), to combine the information from multiple rankers into two lists, respectively denoted by $L^l_k$ and $L^u_k$.
Note that this way of combining the outputs, from multiple rankers before triggering the generative task, 
is different from: a) the FiD (Fusion-in-Decoder) family of approaches~\cite{izacard2020leveraging,hofstatter2023fid} which merge the output from different ranked lists into the context for task-specific tuning of the decoder, and b) the RAG-Fusion~\cite{rackauckas2024evaluating,rackauckas2024rag} family of approaches, which modify input queries with an objective to generate diverse lists of top-documents. In contrast to FiD, our method involves only inference-time computations, and different from RAG-Fusion, the objective is not to increase inter-document diversity but rather to improve the relevance of documents retrieved from each source. 


\para{Z-score for Inter-Source Combination}
  
As the retrieved documents across the two information sources are non-overlapping, it is not possible to apply RRF to compute the expected reciprocal ranks of documents across the document lists $L^l_k$ and $L^u_k$.
Since the problem is similar to that of preference elicitation in a dueling bandits setup \cite{duelng_bandit}, a standard technique is to employ probabilistic comparisons to select the next candidate document from one of the two lists. For these stochastic comparisons, it is a standard practice to assume that the document scores in each list follow a Gaussian distribution \cite{duelng_bandit}. The difference of this problem of inter-source combination of ranked lists with a standard dueling bandit problem is that in our case no rewards are available to improve the selection policy. As such, we simply use the z-score statistic, i.e., standardize the scores of each document in the two lists, and use these scores to induce a total ordering across the two lists.
Formally speaking,
\begin{equation}
L_k =
\argmax_k \{\phi(\overline{\theta_C}(d)): d \in \!\!\!\!\! \bigcup_{C \in \{l, u\}}\!\!\!L^{C}_k\},\,\,
\phi(\overline{\theta_C}(d)) = \frac{\overline{\theta_C}(d) - \mu_C}{\sigma_C},
\label{eq:zscore}
\end{equation}
where $\mu_C$ and $\sigma_C$ are the average and standard deviations of the respective lists, i.e., labeled ($L^l_k$) and unlabeled ($L^u_k$). Intuitively speaking, Equation \ref{eq:zscore}
maps the document scores from the respective sources to a standard normal scale $\pazocal{N}(0,1)$, removing collection-specific bias \cite{cummins2014document,arampatzis2011modeling} enabling a fairer comparison between labeled and unlabeled documents.
%
%

To understand the connection between Equations~\ref{eq:rrf} and~\ref{eq:zscore} and the schematic depicted in Figure~\ref{main_fig}, observe that we first aggregate the ranked lists $L^{C,\theta}_k$ produced by different retrieval models $\theta$ for each source $C \in {l, u}$ (labeled and unlabeled), resulting in two fused lists: $L^l_k$ and $L^u_k$. These two source-specific lists are then further combined in the final stage of the hierarchical fusion process\footnote{Code available at: \url{https://github.com/payelsantra/HF-RAG}}.

\section{Experiment Setup}


Our experiments are conducted to answer the following research questions (RQs):
a) \textbf{RQ-1}: Does combining information from different sources and rankers in a hierarchical manner lead to better out-of-domain generalization?
b) \textbf{RQ-2}: What is the relative contribution of multiple rankers vs. multiple sources in an HF-RAG setup?
c) \textbf{RQ-3}: How strongly does retrieval effectiveness correlate with downstream gains?
d) \textbf{RQ-4}: How sensitive is HF-RAG to its hyper-parameters, i.e., the number of examples in the context?

\para{Datasets}
We conduct our experiments on the fact verification task \cite{petroni2020kilt}, where the objective is to predict if an input claim can be either supported or refuted with evidences retrieved from a collection of documents, or there is not enough information in the collection to do either.
%
%
For our supervised and \icln-based approaches, we use the \textbf{FEVER} training set~\cite{thorne-etal-2018-fact} constituting claim-evidence pairs. As the unlabeled data in U-RAG, we use
the Wikipedia 2018 dump (the underlying document collection for the FEVER dataset with available relevance assessments).
For out-of-domain (OOD) evaluation of models trained on the FEVER dataset we employ the test-splits of the following: a) \textbf{Climate-FEVER}~\cite{diggelmann2020climate,thakur2021beir}, comprising climate-related claims (we removed the `disputed' category to maintain a consistent experiment setup), and b) \textbf{SciFact}~\cite{wadden-etal-2020-fact,thakur2021beir}, comprising scientific claims.
Similar to the FEVER dataset, the claims in both these OOD datasets are also labeled as: `support', `refute', or `not-enough-information'.

\para{Retrievers and Generators}  
We employ the following ranking models in our experiments to retrieve the top-similar candidates either from the FEVER training set (labeled data source), or from the Wikipedia collection (unlabeled data source): a) \textbf{BM25}~\cite{10.1561/1500000019}: a sparse lexical model with prescribed settings of its hyper-parameters, i.e., $(k_1,b) = (1.2, 0.75)$, b) \textbf{Contriever}~\cite{izacard2021unsupervised}: a dense end-to-end bi-encoder model, c) \textbf{ColBERT}~\cite{10.1145/3397271.3401075}: a dense end-to-end late interaction model, and d) \textbf{MonoT5}~\cite{pradeep2021expando}: a retrieve-rerank pipeline based on a cross-encoder model (initial ranker set to BM25). For each IR model, we retrieved the top-$50$ candidates for further processing via the RRF pipeline (Equation \ref{eq:rrf}).

We employ two LLMs of differing scales for the prediction: (a) LLaMA 2.0 (70B)~\cite{llama2-awq,touvron2023llama}, representing a relatively large model, and (b) Mistral (7B)~\cite{mistral7b-awq,jiang2023mistral7b}, a much smaller counterpart.

\para{Methods Investigated} \label{ss:methods}

We compare our proposed method, \hfrag, against both parametric baselines that involve supervised fine-tuning (SFT) and non-parametric RAG-based methods, which may utilize labeled and/or unlabeled data.
Among SFT-based methods, we employ the following:
a) \textbf{\rob}~\cite{long2023adapt} --
a common approach, reported in many studies~\cite{Ke2023ContinualPO,brown2020language,Lee2022ComparativeSO}, involving fine-tuning a standard encoder model RoBERTa 
~\cite{liu2019roberta} on the FEVER training dataset as a 3-way classifier mapping claim-evidence pairs to the labels;
b) \textbf{\lora}~\citep{labruna2024retrieve} -- an LLM decoder model is fine-tuned (specifically, Llama-2-7B~\cite{llama2-7b} for our experiments) as a 3-way classifier on FEVER train claim-evidence pairs via the low-rank domain adaptation (LoRA) technique~\cite{hu2021lora}; and
c) \textbf{CORRECT}~\cite{zhang-lee-2025-correct} -- which first learns an evidence-conditioned prompt embedding by means of noise contrastive loss on the FEVER training set of claim-evidence pairs, and then uses this supervised prompt encoder for few-shot inference with labeled data only (L-RAG).

In addition to the SFT-based methods, we also compare HF-RAG with the following \textbf{non-parametric} RAG-based methods.
\uls
\li \textbf{\iclz} ~\citep{labruna2024retrieve,kojima2022large,li2023classification}: This method predicts the class of a claim (support/refute/not-enough-info) without relying on any additional context (labeled or unlabeled information sources) by leveraging the inherent knowledge stored in an LLM.

\li \textbf{\icln}~\citep{long2023adapt,li2023classification}: A standard in-context learning (ICL) workflow that makes use of the labeled data from the FEVER training data to predict the veracity of a claim.
Out of the four available rankers, we select the one that yields the best performance on the FEVER dev set, which, in our experiment setup, turned out to be Contriever.
Contriever was then employed to retrieve a list of similar claims (with their corresponding labels) from the FEVER training set during inference on the test set.

\li \textbf{\rag}~\cite{lewis2020retrieval, labruna2024retrieve, izacard-grave-2021-leveraging}: This uses the unlabeled data source (Wikipedia collection) for contextual generation via an LLM. Similar to \icln, the ranker model was the best performing one on the FEVER dev set, which turned out to be Contriever for Llama and ColBERT for Mistral. The optimal ranker for a particular LLM was then used to retrieve potentially relevant contextual information from Wikipedia during inference on the test set.

\li \textbf{\icln-RRF}: Instead of applying \icln~ on the optimized ranker, here we apply all rankers to retrieve 4 ranked lists of top-50 candidates, following which, we merge them into a single list by RRF (Equation \ref{eq:rrf} with the labeled data source, i.e., $C=\{l\}$).

\li \textbf{\rag-RRF}: Similar to \icln-RRF, except this uses the unlabeled data source to obtain the 4 different ranked lists, which are then combined via RRF to yield $L^u_k$ (Equation \ref{eq:rrf} with $C=\{u\}$).

\li \textbf{\moe-$\alpha$}: This is an ablation for the z-score based combination strategy - a part of our proposed method HF-RAG. Here, we apply a different strategy to combine the top-lists retrieved from the labeled and the unlabeled sources. Specifically, we use a linear combination (parameterized by $\alpha$) that controls the relative proportion of top-documents to be selected from $L_k^u$ - the remaining (1-$\alpha$) selected from the labeled source, $L_k^l$. A grid search on the FEVER train set was used to optimize $\alpha$.

\li \textbf{RAG-OptSel}: 
This acts as an upper bound on the performance achievable by any single-ranker, single-source RAG configuration selected from the 8 possible combinations in our setup (4 rankers × 2 sources). The best result among these 8 predictions is chosen using ground-truth labels from the corresponding test sets. The goal is to assess whether the proposed combination method can outperform this upper bound.
\ule

\begin{table}[t]
\centering
\small
\begin{adjustbox}{width=0.95\columnwidth}
\begin{tabular}{@{}l rrr rrr@{}}
\toprule
&  \multicolumn{2}{c}{In-Domain} & \multicolumn{4}{c}{Out-Domain}\\
\cmidrule(r){2-3}
\cmidrule(r){4-7}
Predictor & \multicolumn{2}{c}{FEVER} & \multicolumn{2}{c}{Climate-FEVER}& \multicolumn{2}{c}{SciFact}\\
\midrule
\rob  & \multicolumn{2}{c}{0.3010} & \multicolumn{2}{c}{0.2291} & \multicolumn{2}{c}{0.2371} \\
\lora & \multicolumn{2}{c}{0.3959} & \multicolumn{2}{c}{0.3571} & \multicolumn{2}{c}{0.3489} \\
CORRECT & \multicolumn{2}{c}{0.3276} & \multicolumn{2}{c}{0.3295} & \multicolumn{2}{c}{0.3643} \\
\midrule
 & Llama  & Mistral & Llama  & Mistral  & Llama  & Mistral \\
\cmidrule(r){1-7}
\iclz  & 0.4260 & 0.4623 &0.4126 & 0.3724& 0.3297& 0.3258\\
\icln  & 0.4880 & 0.4890 & 0.4602 & 0.3901 & 0.3518 &0.3347 \\
\rag  & 0.4889 & 0.4880& 0.4072 & \underline{0.5083} & 0.3719& 0.4168 \\
\icln-RRF  & 0.5418 & 0.5583 & 0.4755 & 0.4468 &0.3948 & 0.3665 \\
\rag-RRF  & 0.4803 & 0.5185 & 0.4798& \textbf{0.5249} & \underline{0.4012} & 0.3963\\
\moe-$\alpha$  & 0.4880 & 0.3955 & \underline{0.4815} & 0.3703 & 0.3623 & 0.3178\\
\hfrag & \textbf{0.5744} & \textbf{0.5628} & \textbf{0.4838} & 0.5019 & \textbf{0.4320} & \textbf{0.4341} \\
\gc{RAG-OptSel}  & \gc{\underline{0.5468}} & \gc{\underline{0.5584}} & \gc{0.4717} & \gc{0.5001} & \gc{0.3953} & \gc{\underline{0.4246}} \\
\midrule
\end{tabular}
\end{adjustbox}
\caption{
Performance of {\hfrag} relative to the baselines. The best results for a particular experiment setting are bold-faced, and the second-best results are underlined. RAG-OptSel results are grayed out to indicate that it is only a performance bound (using the test labels).
The table reports macro F1 scores, obtained with a context size of 10, i.e., $k=10$ in Equation \ref{eq:zscore}. 
}
\label{tab:cf_combined_results}
\end{table}

\section{Results}
Table~\ref{tab:cf_combined_results} compares our proposed approach and the baselines for in-domain and OOD evaluation.
First, for \textbf{RQ-1} (\textbf{OOD generalization}), we observe that HF-RAG mostly outperforms both parametric and non-parametric baselines
not only for OOD but also for in-domain evaluation. Particularly encouraging are the large improvements observed for scientific claims (SciFact results in Table \ref{tab:cf_combined_results}), as the results show that combining information sources potentially mitigates overfitting a model to a particular domain, e.g., the FEVER model generalizing well for the scientific domain.

In relation to \textbf{RQ-2} (\textbf{multi-rankers vs. multi-sources}), Table \ref{tab:cf_combined_results}
shows that fusion with multiple rankers improves RAG effectiveness with both labeled and unlabeled sources (\textit{L/U-RAG-RRF results, in general, better than L/U-RAG}). Eventually combining information across the two sources further improves results (\textit{HF-RAG results outperforming L/U-RAG-RRF ones}). Combination via z-score is better than the proportional mixture of information from labeled and unlabeled sources (\textit{HF-RAG outperforming LU-RAG-$\alpha$}), which indicates that z-score transformation is able to better capture the relative preference between the documents from the two sources.

\begin{figure}[t]   
\centering
\begin{subfigure}{0.3\columnwidth}
  \centering
  \includegraphics[width=1\textwidth]{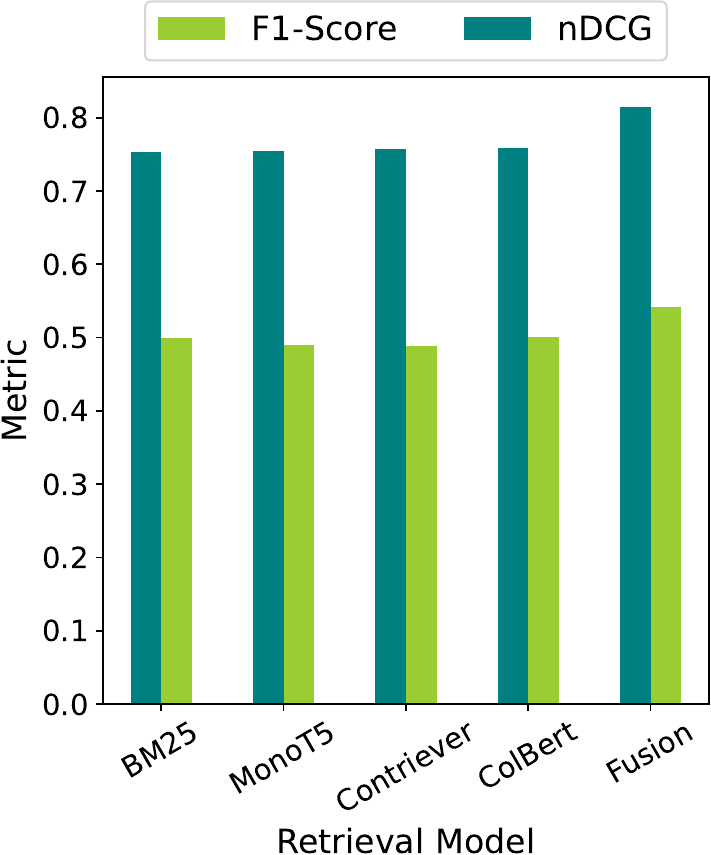}
  \caption{FEVER dataset}
  \label{fig:fev_ndcg}
\end{subfigure}
\begin{subfigure}{0.3\columnwidth}
  \centering
  \includegraphics[width=1\textwidth]{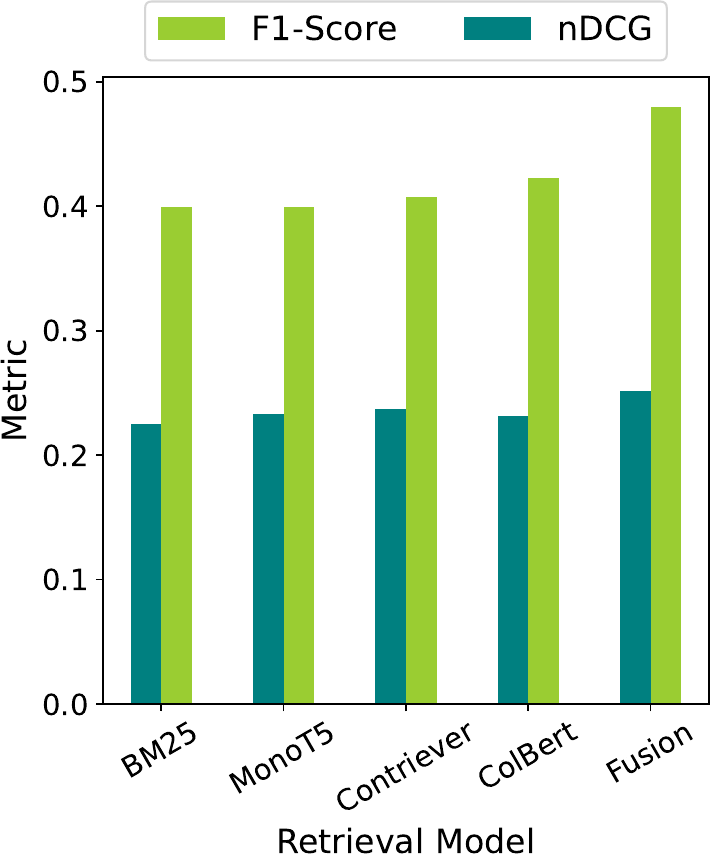}
  \caption{Climate-FEVER}
  \label{fig:c_fever_ndcg}
\end{subfigure}
\begin{subfigure}{0.3\columnwidth}
  \centering
  \includegraphics[width=1\textwidth]{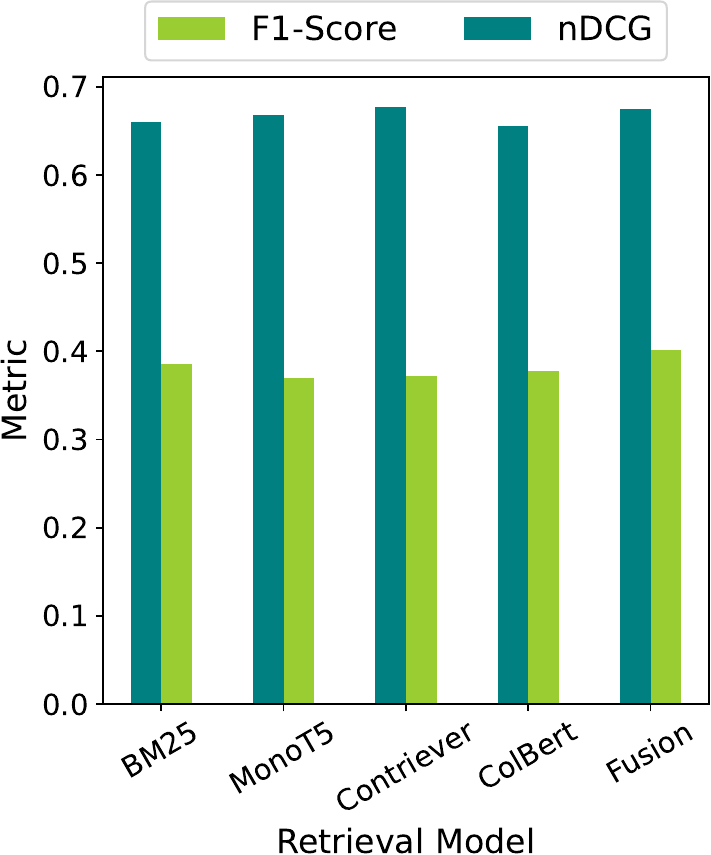}
  \caption{SciFact Dataset}
  \label{fig:scifact_ndcg}
\end{subfigure}
\caption{Comparison between IR (nDCG@10) and claim verification performance (F1) for \rag~ with various models, and \rag-RRF. 
}
\label{fig:ratio_exp_n}
\end{figure}

In relation to \textbf{RQ-3} (\textbf{correlation between retriever and generator performance}), Figure~\ref{fig:ratio_exp_n} demonstrates a positive correlation between retrieval quality--measured by the relevance of evidence retrieved from the unlabeled source--and downstream task performance. The plots indicate that combining multiple rankers consistently improves nDCG@10 across all three datasets. This ranker fusion also results in gains in F1 score, further supporting the benefit of enhanced retrieval quality on end-task performance.

\begin{figure}[t]
\centering
\begin{subfigure}{0.6\columnwidth}
  \centering
  \includegraphics[width=1\textwidth]{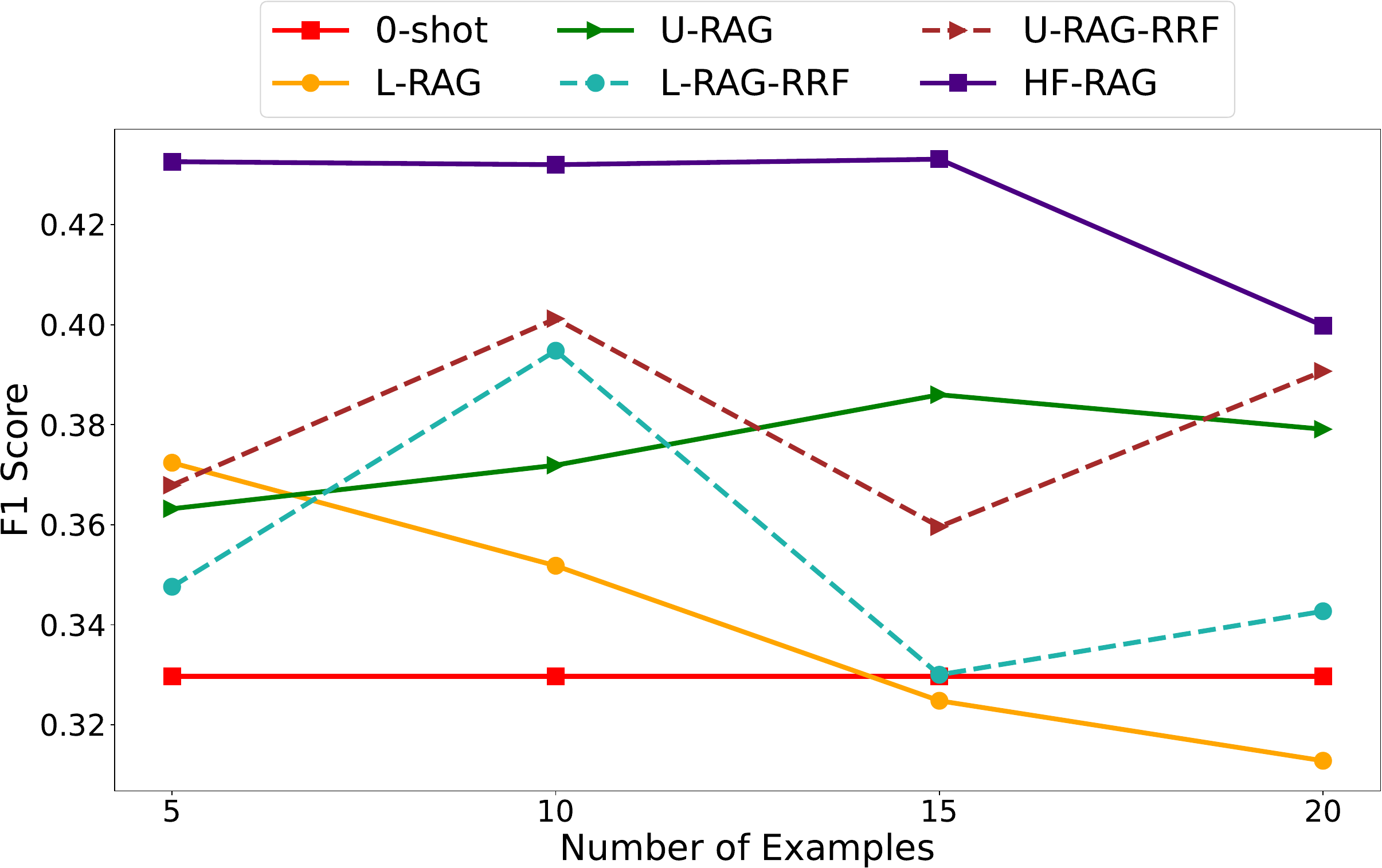}
  \caption{Sensitivity on SciFact}
  \label{fig:js_ndcg_lex_dl3}
\end{subfigure}
\begin{subfigure}{0.32\columnwidth}
  \centering
  \includegraphics[width=1\textwidth]{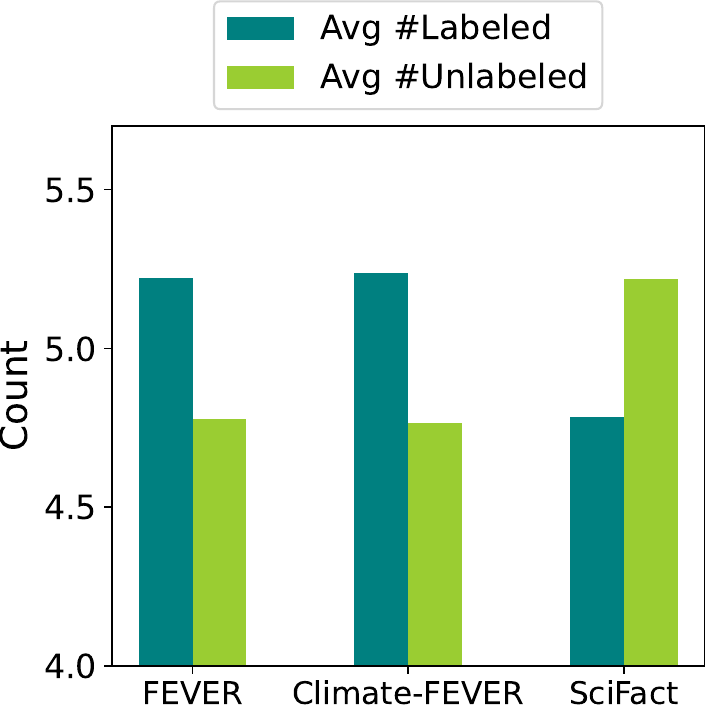}
  \caption{Labeled vs. Unlabeled}
  \label{fig:js_ndcg_clmt}
\end{subfigure}
\caption{(a) Parameter sensitivity of the RAG methods on SciFact predictions; (b) Relative proportion of labeled and unlabeled data in \hfrag~with 10 examples.}
\label{fig:ratio_exp_nw}
\end{figure}

In relation to \textbf{RQ-4} (\textbf{parameter sensitivity of HF-RAG}), we observe from Figure~\ref{fig:js_ndcg_lex_dl3} that HF-RAG exhibits greater stability with respect to context size (i.e., the number of retrieved examples), consistently outperforming both L-RAG and U-RAG as well as their inter-ranker combinations.
Furthermore, in connection with \textbf{RQ-2}, Figure~\ref{fig:js_ndcg_clmt} shows that HF-RAG effectively leverages appropriate proportions of data from labeled and unlabeled sources. Among the two OOD datasets, Climate-FEVER is more similar to FEVER in terms of claim length and linguistic style. In contrast, the scientific claims in SciFact are less aligned with the FEVER domain. Consequently, HF-RAG tends to utilize more information from the labeled dataset—particularly veracity labels of the related claims—for Climate-FEVER. For SciFact, however, it relies more heavily on external knowledge sources, which are likely to be more informative than the veracity labels from FEVER, due to the domain shift.

\section{Concluding Remarks}

We proposed a multi-source multi-ranker RAG approach that first, for each source, combines the top-retrieved documents obtained from multiple ranking models and then combines the information from the two sources of data--labeled and unlabeled--into a merged context for RAG.
Our experiments on the fact verification task demonstrated that our method consistently outperforms several baselines, and also improves over the best RAG performance achievable with an individual ranker or source. Moreover, our method was observed to generalize better on out-of-domain datasets.
In the future, we plan to extend this setup of hierarchical fusion involving multiple sources and multiple rankers to multi-agent RAG with a reasoner component, e.g., search-R1 \cite{jin2025searchr1trainingllmsreason}.

\section*{GenAI Usage Disclosure}
Generative AI tools were not used for core idea generation or experimental design. Its use was limited to minor writing and formatting.

\balance



\end{document}